# Dark-Field X-ray Microscopy for 2D and 3D Imaging of Microstructural Dynamics at the European X-ray Free Electron Laser

Sara J. Irvine[1,2,3], Kento Katagiri[2,3,4], Trygve M. Ræder[5], Ulrike Boesenberg[6], Darshan Chalise[2,3,4], Jade I. Stanton[1,2,3], Dayeeta Pal[2,3,4], Jörg Hallmann[6], Gabriele Ansaldi[6], Felix Brauße[6], Jon H. Eggert[7], Lichao Fang[2,3,4], Eric Folsom[7], Morten Haubro[5], Theodor S. Holstad[5], Anders Madsen[6], Johannes Möller[6], Martin M. Nielsen[4], Henning F. Poulsen[5], Jan-Etienne Pudell[6], Angel Rodriguez-Fernandez[6], Frank Schoofs[8], Frank Seiboth[9], Yifan Wang[2,3,4], Wonhyuk Jo[6], Mohamed Youssef[6], Alexey Zozulya[6], Kristoffer Haldrup[5], and Leora E. Dresselhaus-Marais[2,3,4*]

*[1]Stanford University, Department of Applied Physics, Stanford, 94305, USA*
*[2]SLAC National Accelerator Laboratory, California 94025, USA*
*[3]PULSE Institute, Stanford University, Stanford, 94305, California, USA*
*[4]Stanford University, Department of Materials Science and Engineering, Stanford 94305, USA*
*[5]Technical University of Denmark, Department of Physics, Kgs. Lyngby 2800, Denmark*
*[6]European X-Ray Free-Electron Laser Facility, 22869 Schenefeld, Germany*
*[7]Lawrence Livermore National Laboratory, California 94550, USA*
*[8]FS Consulting, Marlow, United Kingdom*
*[9]Center for X-ray and Nano Science CXNS, Deutsches Elektronon-Synchrotron DESY, 22607 Hamburg, Germany*

*Corresponding author. Email: leoradm@stanford.edu

**Dark field X-ray microscopy (DXFM) can visualize microstructural distortions in bulk crystals. Using the femtosecond X-ray pulses generated by X-ray free-electron lasers (XFEL), DFXM can achieve sub-µm spatial resolution and <100 fs time resolution simultaneously. In this paper, we demonstrate ultrafast DFXM measurements at the European XFEL to visualize an optically-driven longitudinal strain wave propagating through a diamond single crystal. We also present two DFXM scanning modalities that are new to the XFEL sources: spatially 3D and 2D axial-strain scans with sub-µm spatial resolution. With this progress in XFEL-based DFXM, we discuss new opportunities to study multi-timescale spatio-temporal dynamics of microstructures.**

I. Introduction

The ultrafast time scales achievable since X-ray free electron lasers (XFELs) became available have allowed for the study of phenomena previously inaccessible in the fields of condensed-matter physics and materials science. The naosecound to picosecond time scales achievable by XFELs has enabled high-impact advances to the field of *functional materials*. Laser-induced changes in the lattice structure, such as charge density waves [Moore 2016] and changes in magnetic domains [Zusin 2022] have been studied using ultrafast X-ray diffraction (XRD) – offering foundational insights into time scales and mechanisms describing the interactions between electronic structure and material structure. To expand on the insights available from these measurements, certain problems require details of the heterogeneity arising from local structure that define the dynamics and disorder created by these types of changes. Such insights require ultrafast microscopy with diffraction contrast.

Spatial information from diffraction-based imaging provides insight into local structural variation not achievable by XRD measurements alone. Diffraction-contrast imaging explored at the XFEL, includes techniques such as Coherent Diffraction Imaging (CDI) [Clack 2012, Choi 2023], dark-field X-ray microscopy (DFXM) [Dresselhaus-Marais 2023] and topography [Rodriguez-Fernandez, 2021]. Each technique has unique trade-offs for understanding particular problems. DFXM provides insight into bulk samples which would be of specific interest in materials in which small amounts of surface strain would dramatically change the material properties. DFXM offers a complimentary set of tools, which has the potential to provide important insight in the study of quantum materials.

DFXM uses an X-ray objective lens to produce images from the X-ray diffracted beam [Simons, 2015]. An X-ray objective lens then magnifies the diffracted beam to form a real-space image on a camera in the image plane. The use of Compound Refractive Lenses (CRLs) [Snigirev, 1996] enables versatile focusing power for high photon-energy imaging with relatively high spatial resolution, reaching 150-nm using synchrotron X-rays [Kutsal, 2019]. Similar to X-ray topography [Gaukroger 2008, Moore 2009], full-field DFXM images spatially map the deviations in crystal packing (tilt and strain) of a specific crystallographic plane [Poulsen 2017,2018,2021,has strong application to explore novel phase transitions in quantum materials, specifically at the ultrafast timescales available at the XFEL. This work further develops DFXM at the XFEL, for applications that apply to metallurgy but also to the study of ultrafast phenomena in *functional materials.*

Use of the XFEL-based DFXM has a strong advantage to study ultrafast dynamics that relate to lattice displacements. Holstad et al. were the first to use an XFEL to perform DFXM measurements to record a movie of the strain waves propagating inside a single-crystalline diamond [Holstad 2023]. Now that XFEL-based DFXM has been demonstrated, many of the scanning modalities available with synchrotron-based DFXM [Brennan 2022, Yildirim 2023, Huang 2023] must be extended to the unique aspects of the XFEL setup and beam characteristics [Holstad 2022]. In this work, we expand the range of DFXM scans available at XFELs with the first demonstration of the technique at the Materials Imaging Dynamics (MID) instrument of the European XFEL [Altarelli 2007, Tschentscher 2017, Madsen 2021]. Our DFXM measurements of 3D structural and axial strain scans visualized an ultrafast longitudinal strain wave in {111} oriented diamonds demonstrating the new information these scanning methods offer to those published previously. In the sections below, we detail the raw data and 2D rocking curves consistent with the previous works, then include a spatially 3D dataset of our microstructures and an axial strain scan of the longitudinal wave. We then discuss how these scans and our higher magnification in this work extend the XFEL-based DFXM measurements, including a discussion of some of the ultrafast science that is now feasible for future XFEL-based DFXM experiments.

II. **Experimental methods**

The experiment in this work was carried out on a custom-built DFXM setup that was installed at the MID instrument of the European XFEL, using a pump-probe configuration with fs-duration pump laser [Madsen, 2021]. Fig. 1a displays a schematic of our experimental geometry, and we include an example of a typical DFXM image of single-crystalline diamond obtained in this work (Fig. 1d). The diamond sample used in this work was synthesized via the high pressure and high temperature method that was cut and polished along (111) planes with a ~3° offset to avoid cracking (used as received from Element Six, Inc.). An aluminum layer with a thickness of 100–200 nm was sputter deposited onto the upstream (laser-irradiated) side of the diamond to serve as a transducer material to generate the strain pulse via the photothermal effect [Pezeril, 2016]. The nominal XFEL photon energy was 10.020 keV (except when *E* was scanned explicitly), and used a self-seeded monochromatic pulse [Liu et al 2023,] to achieve a 0.1% bandwidth ($\Delta E/E$) at the highest possible photon flux.

To produce the DFXM images, the XFEL beam was pre-focused into a vertical line-beam using a stack of 8 1D diamond CRLs with 100-µm radius of curvature and 1-mm thickness (PALM Scientific LLC), using 400-µm diameter circular pinholes at both the entrance and exit of the stack. The sample was placed 83 cm from the exit-surface of this 1D CRL stack, resulting in a calculated spot of 400 µm tall and 1.7 µm wide at the sample. The incident vertical-line beam at the sample illuminated an *observation plane* that spans the depth of the diamond sample, defining the volume that was diffracted along the diamond (111) peak in the horizontal plane of the lab. The diffracted beam was then imaged and magnified with a stack of 37 beryllium 2D CRLs with 0.5-mm thickness and 50-µm curvature radius. We placed a 300-µm CuAlNi aperture at the exit surface of the beryllium CRL stack to remove stray scatter. The relevant imaging distance between the sample and the entrance aperture of the 2D imaging CRLs was 22.3 cm ($d_1$), and the distance from the exit-aperture to the detector screen was 9.3-m ($d_2$), offering a total X-ray magnification of 17.3x. [Simons 2017]The X-ray image was then detected using an indirect detection scheme, in which the image was converted to visible light at a 0.02-mm thick Ce:YAG scintillator screen, then further magnified by a 10x infinity corrected Mitutoyo objective and recorded by the 6-µm/pixel Andor Zyla 5.5 sCMOS camera. The resulting 173x total magnification had a measured spatial resolution of 0.74-µm. Calibration of the spatial resolution and camera focus were performed using a 400-mesh copper TEM grid at the rear surface of a diamond sample and in front of the scintillator, respectively.

The strain wave studied in this work was launched by a 50-fs pulse of $\lambda = 800$ nm following the pump-probe experiment used in [Holstad, 2023] and described in full in [Pergament, 2015]. The optical pump laser was aligned to be nearly collinear with the XFEL beam, and focused to a 2D Gaussian spot with a 180-µm diameter (FWHM) and estimated pulse energy of 22 µJ at the sample (geometry shown in Fig. 1b). The diamond sample was rotated by 2.3° with respect to normal incidence of the XFEL beam to satisfy the diffraction condition (Fig. 1c). The laser irradiated the aluminum layer to generate a transient thermal expansion pulse that was then transduced into the diamond. While some complex structure is generated by acoustic waves produced in this way \cite{Darshan,2024}, our DFXM in this work only captured the primary strain wave for our selected diffraction conditions. The XFEL and pump laser spot sizes are shown in Fig. 1b.

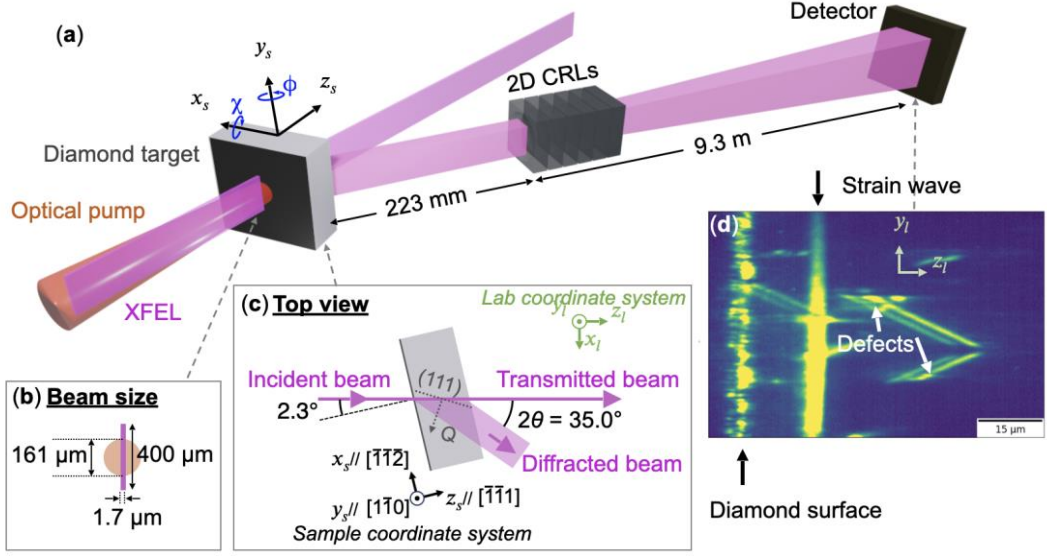

**FIG 1.** (a) Schematic of the *in-situ* DFXM experiment performed at MID, European XFEL (not drawn to scale), with (b) an overlay of the beam sizes for the XFEL and visible laser. (c) The laboratory coordinate system ($x_l$, $y_l$, $z_l$) and the sample coordinate system ($x_s$, $y_s$, $z_s$) are defined as shown. The $z_s$ is normal to the diamond surface which is oriented along the (-1-11) plane to appropriately define diffraction from the Q=(111) plane. (d) A typical single-shot DFXM image collected for a weak-beam condition, visualizing the strained diamond surface, laser-driven strain wave, and the pre-existing defects inside the diamond.

Following a similar geometry to the one outlined in [Dresselhaus-Marais, 2023], sample alignment used three rotation stages and four translation stages, with duplication of the *y*-stages ($y_1$ and $y_2$). As was done in that work, we define our translation stages in a lab coordinate system aligned to the incident beam, ($x_l$, $y_l$, $z_l$), and the three rotation axes of the $\chi$, $\varphi$, are set along $x_l$, and $y_l$, respectively (Fig. 1a), before any rotations to define the goniometer. The stacking order of these stages (bottom to top) are: $z$, $y_1$, $x$, $y_2$, $\chi$ then $\varphi$ with the $z$-stage being closest to the sample. We note that after rotations begin, the goniometer stages are no longer aligned to the lab system, instead defining the rotational transforms to set the sample system ($x_s, y_s, z_s$). During alignment, the three rotational stages are used to identify the crystal orientation with the most intense diffraction signal.

Once the crystal was aligned into the most intense Bragg condition, rotational scans (rocking curve) were performed to resolve the spatial extent of different directions of lattice displacement along the (111) plane. The setting where the observed Bragg peak intensity gets maximized is termed as the *strong-beam condition*, and is used as the reference orientations ($\mu_0$, $\varphi_0$, $\chi_0$) for subsequent scans [Jakobsen 2019]. The sample is scanned individually along $\varphi$, $x_s$, and the XFEL photon energy, to perform the rocking-curve scan, 3D structure scan, and axial strain scan as fully described in the Results Section.

**III.  Results**

For all DFXM images collected in this work, dark frames were collected regularly and were subtracted from the signal images to correct the background and enhance the signal-to-noise. The process for image processing is further discussed in Supplementary Section II.

**A. Rocking-Curve Scan**

In this section, we begin with the one-axis rocking-curve scans performed for our DFXM studies – building on similar ones performed routinely at synchrotrons [Ferrer 2023] and in the previous XFEL-DFXM study [Dresselhaus-Marais 2023]. During the rocking curve ($\varphi$) scan, the sample is rotated through the Bragg condition, capturing 20-50 images at rotational steps along $\varphi$ that span a range beyond the edges of the diffracted signal on either side of the diffraction peak. In these single crystalline samples, the *strong-beam condition* defines the image collected at the center of the Bragg peak, while each side of the rocking curve produces images in a *weak-beam condition* that produces bright image features from only regions of a specific strain state in the crystal arising from defects, strain waves, surface strains, etc.

The center $\varphi$ value for a given rocking curve scan (*i.e.*, the strong-beam condition) is determined by fitting the mean intensity of each image to a Gaussian curve. In this and previous DFXM studies, the $\varphi$ position is defined as the angular deviation from the center of the rocking curve. The low divergence of the collimated XFEL beam makes DFXM's resolution to strain of apprxaility an order of magnitude at XFELs than is observed at synchrotrons [Holstad, 2022]. As such, the resolution of a $\varphi$ scan at an XFEL is mainly limited by the precision and step size of the rotation stage; in this work, the result was a$\varphi$ scan was 0.003° in this work.

Selected DFXM images are shown in Fig. 2 for the three characteristic $\varphi$ positions along a $\varphi$-rocking curve scan collected at the nominal $\chi_0$ position. The two images showing the weak beam conditions for $\Delta\varphi = -0.013°$ (Fig. 2a) and $\Delta\varphi = 0.014°$ (Fig. 2c). This rocking scan was collected for different $\varphi$ rotations of the crystal for a constant pump-probe delay time of $\tau = 2.0$ ns, and thus show the different lattice rotational states present in the crystal at that point in the strain wave's propagation. The $\varphi$ position in which a given pixel's signal is maximized corresponds to the symmetry state that describes the maximum population of the lattice strain/rotation at that point. As the scan was collected in discrete $\varphi$ steps without continuously rotating between over each $\Delta\varphi$ range, not all strain/rotational steps were measured. As no image was not collected at the ideal strong beam condition, a fit was used to select the $\varphi = 0$ position defining the center of the scan, introducing an offset of $\Delta\varphi = 0.002°$ (Fig. 2b) from the scan's resolution.

The observed differences between the images obtained at two opposite weak beam conditions

(Fig. 2a,c) arise from the difference of the positive and negative strains along the probed symmetry axis. We also note that the strong-beam condition (Fig. 2b) displays fringe patterns that are characteristic of interference caused by dynamical diffraction. The coherence of the XFEL and long mean-free-path for scattering in low defect-density diamond gives rise to significant dynamical effects, which sometimes appear as interference patterns under the appropriate conditions. A discussion of the nature of this dynamical diffraction is included in the Discussion Section.

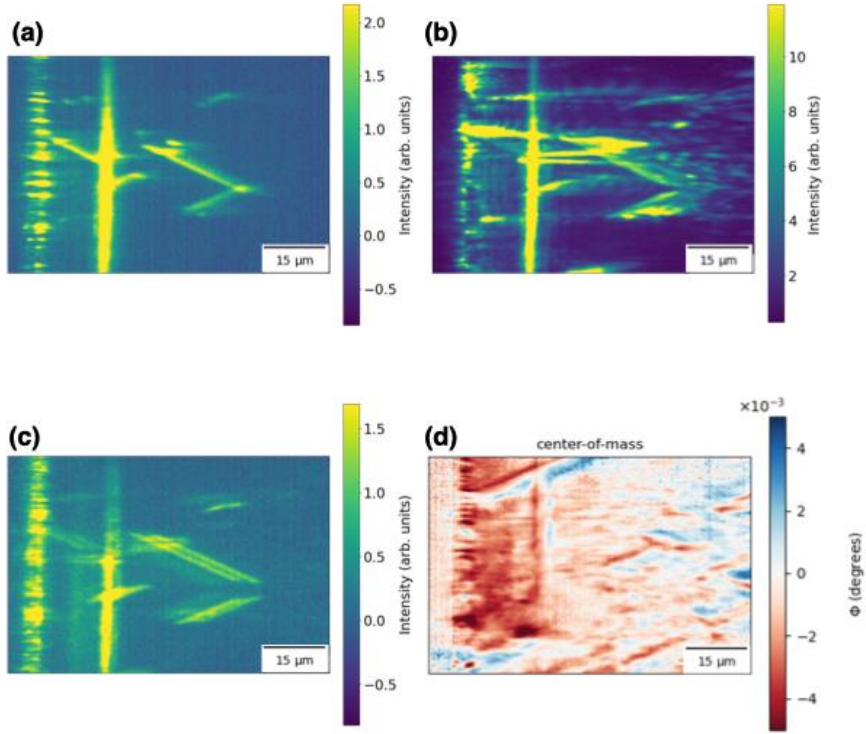

**FIG 2.** DFXM images obtained from a rocking-curve scan, with the crystal oriented along a $\chi_0$ value that passes through the highest-intensity diffraction condition. Raw images from three different $\varphi$ positions are shown for (a) a weak beam condition at $\varphi = -0.013°$, (b) the strong-beam condition at $\varphi = +0.002°$, and (c) the other weak beam condition at $\varphi = +0.014°$. (d) The center-of-mass (COM) image compiled from a rocking curve scan; the COM image displayed in (d) is from a different region of the sample than the images in (a)-(c) for feature clarity.

A full stack of images like those in Fig. 2a-c were collected over the selected range of $\varphi$ for the full rocking curve, comprising 21 separate $\varphi$-images each describing an average of ~200 frames that integrated XFEL pulses for 20 seconds at each unique $\varphi$ position. The integrated image series for the $\varphi$-scan was then processed into a center-of-mass (COM) plot like the one shown in Fig. 2d to convey the regions most representative of each strain/rotation state in the sample. The color on each pixel in the COM plots describes the $\varphi$ position over the full stack for which the maximum intensity value, $I(\varphi)$, was recorded for diffraction at that pixel. As described in [Ferrer 2023], the maximum $I(\varphi)$ is identified as the center of a Gaussian fit of the $I(\varphi)$ values. The COM plot shown in Fig. 2d is from a different $\varphi$-scan from the scan included in Fig 2a-c. The sample region shown in Fig. 2a-c shows dynamical diffraction strongly, but we selected a region of the sample that had minimal dynamical diffraction effects in Fig. 2d to display the COM plot to avoid artifacts in the reconstruction. The COM scan in Fig. 2d visualizes the local strain along over the field of view within the observation plane. The wider number of strain states represented in the COM images enables image features to be identified such as the dislocations, the strain wave, and the interface connecting the diamond's surface to the aluminum transducer. In materials that include more intricate patterns in the microstructure, the strain contrast with the COM is more important to discriminate between their characters (e.g. shear vs longitudinal acoustic waves).

**B. 3D Structure Scan**

We note that the scans in Fig. 2 all describe a single observation plane, without 3D information. To understand the out-of-plane information about the microstructure requires further scanning of the sample, as shown in [Yildirim 2023]. Here we demonstrate the first 3D ($x$, $y$, $z$) scan of the XFEL-based DFXM which has promising potential to resolve ultrafast 3D dynamics with the highest spatial and temporal resolutions. In this scan, we repeated the above rocking curve ($\varphi$) scans at 40 different $x$ positions and associated ($y$-$z$) observation planes, spaced 4-µm apart. The resulting 4D dataset ($x$, $y$, $z$, $\varphi$) was then visualized by downselecting each $\varphi$-scan into a single representative image that defines the weak-beam condition that best displays the defects in that layer. Due to the high sensitivity of this technique, it is not sufficient to choose a single $\varphi$ position for each $x$-defined plane (both due to motor precision and gradients in material structures). As such, the representative weak-beam frames were identified using a Gaussian fitting routine with a few manual adjustments, as described in Supplementary Section IV. A typical result of the 3D structure scans is shown in Fig. 3, which includes dislocations and stacking faults in the diamond.

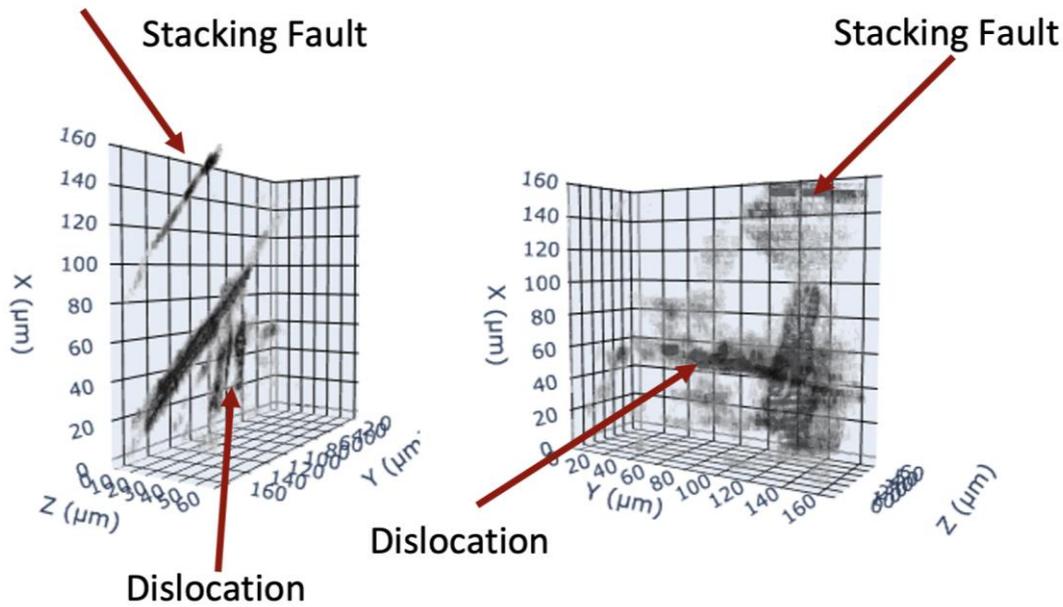

**FIG 3.** The 3D microstructure inside the single-crystalline diamond measured by DFXM in this work. The diamond surface is at the *x*-plane at *z* = 0 but is omitted for image clarity. The black planar and linear features represent stacking faults and dislocations inside the diamond, respectively, in the diamond.

The analysis previously performed for data collected from the European Synchrotron Radiation Facility [Yildirim, 2023] was modified for this work to match the unique attributes of the XFEL dataset. While the synchrotron case used continuous scanning of both $\phi$ and *x* throughout the image acquisition, the pulsed modes of the XFEL could only measure discrete steps along the scan — meaning that some of the reciprocal-space information is under-sampled. With larger step sizes in both the $\phi$ and *x* directions motors, this led to additional challenges in interpolation and integration of the datasets.With larger step sizes in both the $\varphi$ and *x* directions, this led to additional challenges in interpolation and integration of the datasets.

Due to the linear defect features in diamond an the limited sampling capabilities of the xfel compared to the synchrotron where this scan method was established, different processing methods were used in comparison to previous work. [Yildrim 2023]. Due to the fact that the observed features were limlar and the largest noise source was seemingly hot pixels. A median filter was used with a kernel that was off set with a smaller value in the x direction. The acquired images were filtered by using a standard image filtering process involving 5 loops of erosion and dilation [González, 1993] to reduce the main cause of noise in this case arising from hot pixels at the detector. Since the observed defects are linear or planar, it was possible to remove this noise without degrading the observed defect features. The processed data is then binned by 5 x 5 to reduce computational costs and plotted with the marching cubes algorithm [Lorensen 1998] with scaling relevant to the 3D spatial axes from the scan.

**C. Axial Strain Scan**

At synchrotrons, DFXM can measure axial strain along the diffraction vector using a $\theta$-$2\theta$ scan; this requires synchronized rotation of the crystal, lens, and detector along the same concentric arc [Simons 2018]. Accurate synchronized motion of these stages becomes prohibitively challenging at XFELs as, unlike synchrotrons, no XFELs currently have dedicated DFXM facilities that can be globally calibrated for routine scanning. Here we demonstrate an alternate approach for axial strain scans that is simpler to implement at XFELs: incrementally scanning the X-ray photon energy to sample the strain at a constant $2\theta$. For this scan, we held the original alignment of the detector and lens at the $2\theta$ value for the strong-beam condition for the original photon energy (the center of our energy scan). We then changed the XFEL energy to change the effective $d$-spacing that meets the Bragg condition for that $2\theta$ position.

While a $\varphi$ scan alone is not sufficient to efficiently decouple rotational (shear) vs axial strains, by scanning both the photon energy and $\varphi$, this may become possible [Poulsen 2021]. The accelerator modes for *on-demand* photon energy tuning of the seeded-monochromatic XFEL beam are still relatively nascent [Liu et al. 2023], however, using data points taken at 4 different photon energies (10.011, 10.017, 10.020, and 10.026 keV), our axial scan of the (111) peak demonstrates the viability of energy scans to decouple the axial and rotational strains in crystals at the ultrafast timescales only accessible at XFELs. Small steps of 6-eV or 3-eV were selected to be within the anticipated rocking curve due to the sensitivity of the microscope to $\theta$. As the translation stages for the microscope optics/detectors were not changed at each photon energy, the $d_1$ and $d_2$ imaging distances and angles were the ones optimized for 10.020 keV, introducing small chromatic effects at the rocking-curve edges. For a full summary of the scan statistics and chromatic effects, we refer the reader to Supplementary Section III.

Fig. 4 includes the COM reconstructions for a representative $\varphi$ rocking curve collected at 10.020 keV (Fig. 4a) for comparison to the axial strain COM scan obtained from our abbreviated photon energy scan (Fig. 4b). The limited number of the energy steps used to create the COM plot for the energy scan causes an artificially low signal-to-noise ratio because the scan is under-sampled in reciprocal space. Since reciprocal space maps the 3D momentum states defining the lattice, fully sampling the space requires free motion along the three principal axes. The color bar is also off set as since 10020eV was the energy the experiment was aligned for, there were two data points collected at a lower energy and only one at a higher. In the synchrotron goniometers for DFXM, this is spanned by $\varphi$, $\chi$, and $2\Delta\theta$. In the present setup, the $\varphi$-$E$ scan contains both rotational and axial strain information because the position of $\chi$ was not optimized between each photon energy. As such, comparison between Fig. 4a and b is helpful to offer context to the coupling between the axial and rotational strain in features from each image.

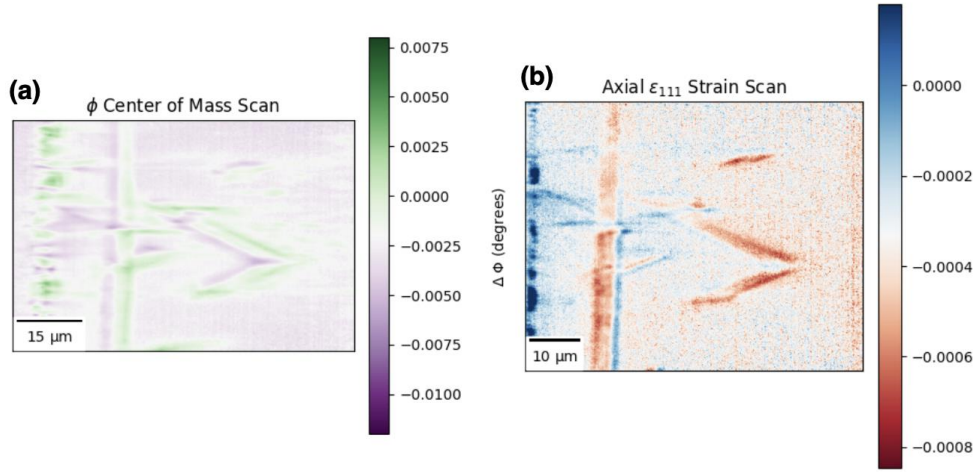

**FIG 4.** Center of Mass (COM) plots for the (**a**) rotational and (**b**) axial strain scan of the same region of the sample, illustrating the complementary information contained in these views.

**IV. Discussion**

As discussed above, DFXM has been developed as a dedicated setup at a few synchrotrons, where a wide range of scanning methodologies have been established. Permanence in a DFXM setup enables more controlled development of scanning methodologies, but also because of the design choices for the setup. At a synchrotron, the goniometer motors are selected with careful consideration of the stability and accuracy with respect to the unique sensitivity of DFXM. By contrast, the *ad-hoc* nature of XFEL experiments tends to prefer well-calibrated stages that are calibrated for a versatile set of experiments. As such, the extension of DFXM scanning methods to the XFEL has required computational techniques or technical approaches to measure the same sample observables. For example, absolute motor calibration at synchrotrons allows for multi-axis scans like the lens holder + detector scans along 2θ. At the XFEL, those same motor calibrations are not viable and thus an energy scan that avoided motorized translation was more accurate. In this work, we demonstrate the XFEL versions of a DFXM rocking scan, 3D structure scan and strain scan, offering access to ultrafast timescales.

The importance of the implementation of the 3D structure scan at the XFEL can be seen in a more rigorous characterization of the defect structure in diamond. The 2D images shown in Fig. 2 include numerous linear distortions in our diamond single crystals. It is easy to assume that the linear features are dislocations (line defects); however, since DFXM only observes a planar region through the depth of the crystal, the linear features could alternately correspond to planar structures slicing through the observation plane. Indeed diamond is known to form stacking faults, which form planar structures inside the crystal [Gaukroger 2008, Masuya 2017]. Our 3D reconstructed image shown in Fig. 3 validated that many of the linear features observed in our 2D images are actually planar stacking faults, but also allowed us to resolve a few dislocations (shown with the red arrow in Fig. 3). This type of detailed structural characterization is essential to distinguish between planar versus linear defects, and will also be important when interpreting the ultrafast structural dynamics.

The images in Fig. 4 also demonstrate the first report of axial strain maps along the diffraction vector at the XFEL with DFXM scans. The rotational and axial strain information about the surface and the internal defects inside the diamond offered an opportunity to discern which strain fields exhibit the most *shear* contributions. Since all components of the strain field have non-zero projections onto the (111) plane of crystals, DFXM's signal in this type of scan could include contributions from all components of the strain tensor field[cite dayeeta]. In our measured data (Fig. 4), we observe intensity changes in both the positive and negative axial strain scan surrounding the strain wave. By contrast, we see only negative strains at the surface. We note, however, that the surface strains may not be representative of the material because the field of view shift with photon energy may have clipped the region with positive strain. As such, we focus our analysis on the strain wave portion of the image. The defect marked by the yellow arrows in Fig. 4 demonstrates asymmetry that emerges in the axial strain, though it has symmetric positive and negative strains in shear. While this type of information offers insights relevant for comparison to models, we omit full discussion of the asymmetry in strain vs mosaicity due to the limited sampling in this preliminary axial strain scan. The limited sampling was caused by the 1-hr times required to acquire data at each strain position, making more than 5 data points in the scan unviable in the limited experimental time. This long duration was caused by the longer time required to change energy with a seeded XFEL compared to the SASE, in which the bandwidth would have been ~20eV, making this particular experiment faster, though DFXM at an XFEL is greatly enhanced by having a self-seeded beam.[Liu 2023]

As may be observed in Fig. 2, the images collected in our strong beam conditions included complex features arising from dynamical diffraction effects in the image. These dynamical effects appear in the image as fringes that emerge from the linear features of stacking faults, dislocations, and the strain wave in the crystal. We observed fringing effects that we attribute to the dynamical effects caused by the strain fields emerging from the defects and distortions present in an otherwise perfect crystal [Kato et al 1966]. Features of dynamical diffraction have been observed in X-ray topography as shadows or fringes that emerge categorized based on the absorption condition at which different types of excitation contrasts can dominate [Raghothamachar et all 2006]. Some higher-order features in topography images appear as fringes of the *intermediate image* absorption condition – below the dynamical limit at which the feature appears as a single shadow [Raghothamachar et all 2006]. We interpret that this intermediate image extinction contrast is similar to the extinction contrast giving rise to the fringes we observe in DFXM. Such features in topography measurements have given deep insights into sub resolution features in the sample [Gradwoh 2020]. While a detailed analysis of the <u>differences</u> in image features arising from dynamical diffraction between X-ray topography and DFXM are beyond the scope of this work, we note that these effects are amplified by the collimated beams available at XFELs. Future studies on these dynamical diffraction effects offer unique opportunities in materials with the appropriate wave-optics formalism [Carlsen 2022].

Finally, we note that the ultrafast dynamics of the strain wave and other ultrafast features are newly accessible with DFXM at the XFEL. In this and previous works, we have discussed the work we've referred to the longitudinal strain wave with a , and focused on the purely strain effects due to the transduced wave. At timescales accessible at the enable by XFELs, however, the acoustic wave formalism is not the only way to consider the waveform. We also note that, in the future, DFXM at XFEL's can be used to study phonon propagation in materials utilizing the longitudinal phonons created during the ultrafast excitation of the Al transducer. Works by Trigo et. al [Trigo 2013; Zhu 2015] demonstrated that time-resolved measurement of thermal diffuse scattering of X-rays from squeezed phonons can be used to reconstruct phonon band structure in single crystals. The ability to create coherent phonons and perform time-resolved measurements at XFELs can be combined with excellent resolution in real and reciprocal space [Poulen et al, 2021] can enable study of phonon propagation in ways that have not been possible otherwise.

The use of the XFEL for these measurements now enables the ability of DFXM's unique sampling of material structure to study ultrafast phenomena in materials. With the new scanning modalities introduced in this work, the DFXM datasets generated at XFELs can now access the full range of data now accessible at synchrotrons. This now opens opportunities to image ultrafast dynamics with high resolution in *both* real and reciprocal space. Using these approaches, DFXM has potential applications to image the dynamics in 3D axial strain fields generated by pump-probe experiments to drive magnetic domain-wall motion [Takeuchi 2022], transitions to slightly distorted "hidden phases" [Han, 2015], and to study the photoelastic effects generated when light absorption drives a change to the strain fields, e.g. in piezoelectrics [Kim, 2022].

**V. Summary**

This study demonstrated the first implementation of DFXM at the MID instrument of the European XFEL and illustrated three scanning modalities with this new setup. With the addition of an optical microscope objective, our images offered 173x magnification and a spatial resolution of 0.74µm over a 78µm x 93µm field of view. We demonstrated how, with the use of the seeded monochromatic beam, we could achieve high signal-to-noise even at the higher magnification than our previous study. Our results first show representative images of the strong-beam and weak-beam images relevant to a single rocking-curve scan, with a COM plot illustrating their relationship. We then introduce the first 3D DFXM scans from XFELs that we use to convey which defects arose from dislocations vs stacking faults. Finally, we demonstrated the first DFXM experiments to scan the photon energy to collect images of the axial strain to discriminate between shear and longitudinal strains. Our discussion explains how analysis of these results from our pump-probe experiment offers unique insights into the future of ultrafast DFXM. These new scans and DFXM capabilities with the MHz frame rates available at the European XFEL now offer unique opportunities to study ultrafast phenomena at microseconds and faster timescales.

**Supplementary Material**

This supplementary material details the code and image processing used in the study, including experimental geometry, energy scan stability, and a refined method for selecting weak-beam positions. It explains the methodology for 3D structure scans, focusing on optimizing weak-beam frames through Gaussian fits and manual adjustments. The code and data are available via GitHub and a data repository.

**Acknowledgments**

We acknowledge European XFEL in Schenefeld, Germany, for the provision of X-ray free-electron laser beamtime at Scientific Instrument MID (Materials Imaging and Dynamics) and would like to thank the staff for their assistance. Part of this work was performed at the Stanford Nano Shared Facilities (SNSF), supported by the National Science Foundation under award ECCS-1542152.

Stanford/SLAC work from this study was supported by the Department of Energy, Office of Science, Basic Energy Sciences, Materials Sciences and Engineering Division, under Contract DE-AC02-76SF00515. Financial support was also provided by the Villum FONDEN (grant no. 00028346) and the European Spallation Source lighthouse on hard materials in 3D, SOLID, funded by the Danish Agency for Science and Higher Education (grant number 8144-00002B). Moreover, H.F.P. and H.S. acknowledge the support from the European Research Council (Advanced grant no 885022 and Starting grant no 804665, respectively). The authors affiliated with the Technical University of Denmark gratefully acknowledge DANSCATT for supporting the beamtime efforts. T.M.R. acknowledges funding from the European Union's Horizon 2020 research and innovation programme under the Marie Skłodowska-Curie grant agreement No 899987. JE, ENF and BK acknowledge their work under the auspices of the US Department of Energy by Lawrence Livermore National Laboratory under contract DE-AC52-07NA27344.

**AUTHOR DECLARATIONS**

**Conflict of Interest**

The authors have no conflicts to disclose.

**Author Contributions**

**Sara J. Irvine:** Formal analysis (lead); Visualization (equal); Writing – original draft (equal).

**Trygve M. Ræder:** Investigation (equal); Formal analysis (equal); Visualization (equal); Methodology (equal).

**Kento Katagiri:** Investigation (equal); Formal analysis (equal); Visualization (equal); Writing – original draft (equal).

**Darshan Chalise:** Formal analysis (equal); Visualization (equal); Writing – original draft (equal).

**Gabriele Ansaldi:** Investigation (equal).

**Ulrike Boesenberg:** Investigation (equal).

**Felix Brauße:** Investigation (equal).

**Jon Eggert:** Conceptualization (equal).

**Lichao Fang:** Formal analysis (equal).

**Eric Folsom:** Resources (equal).

**Jörg Hallmann:** Investigation (equal).

**Morten Haubro:** Investigation (equal); Visualization (equal).

**Theodor S. Holstad:** Investigation (equal); Visualization (equal).

**Bernard Kozioziemski:** Conceptualization (equal).

**Anders Madsen:** Investigation (equal).

**Johannes Möller:** Investigation (equal).

**Martin M. Nielsen:** Conceptualization (equal); Methodology (equal); Investigation (equal).

**Dayeeta Pal:** Formal analysis (equal).

**Henning F. Poulsen:** Conceptualization (equal); Supervision (equal); Methodology (equal); Investigation (equal); Funding acquisition (lead).

**Jan-Etienne Pudell:** Investigation (equal).

**Angel Rodriguez-Fernandez:** Investigation (equal).

**Frank Seiboth:** Resources (equal).

**Frank Schoofs:** Resources (equal).

**Jade I. Stanton:** Formal analysis (equal).

**Yifan Wang:** Formal analysis (equal).

**Jo Wonhyuk:** Investigation (equal).

**Mohamed Youssef:** Investigation (equal).

**Alexey Zozulya:** Investigation (equal).

**Kristoffer Haldrup:** Conceptualization (equal); Methodology (equal); Investigation (equal).

**Leora E. Dresselhaus-Marais:** Conceptualization (equal); Supervision (equal); Methodology (equal); Investigation (equal); Formal analysis (equal); Writing – review & editing (equal).

## DATA AVAILABILITY

All code used to process the data in this work is available at the Github repository that may be accessed at https://github.com/leoradm/EuXFEL-DFXM-2024, and all data used to construct the figures in this work are available at the data repository https://osf.io/9g36u/?view_only=c5000015c43f4c78bd4a712fe1b0f84f.

## References

1) Altarelli, M.; Brinkmann, R.; Chergui, M.; Decking, W.; Dobson, B.; Düsterer, S.; Grubel, G.; Graeff, W.; Graafsma, H.; Hajdu, J.; Marangos, J.; Pflüger, J.; Redlin, H.; Riley, D.; Robinson, I.; Rossbach, J.; Schwarz, A.; Tiedtke, K.; Tschentscher, T.; Vartaniants, I.; Wabnitz, H.; Weise, H.; Wichmann, R.; Witte, K.; Wolf, A.; Wulff, M.; Yurkov, M. "XFEL: The European X-ray free-electron laser: Technical design report" Report No. DESY 2006-097 (DESY, 2006).

2) Carlsen, M.; Simons, H. "A Finite Difference Scheme for Integrating the Takagi–Taupin Equations on an Arbitrary Orthogonal Grid." Acta Crystallogr A Found Adv, 78 (5), (2022).

3) Chalise D.; Wang Y.; Trigo M.; Dresselhaus-Marais L.E. "Imaging the dynamics of coherent and incoherent phonons with dark-field x-ray microscopy using kinematic diffraction theory" In Preparation (2024).

4) Choi, S.; Im, S. W.; Huh, J.-H.; Kim, S.; Kim, J.; Lim, Y.-C.; Kim, R. M.; Han, J. H.; Kim, H.; Sprung, M.; Lee, S. Y.; Cha, W.; Harder, R.; Lee, S.; Nam, K. T.; Kim, H." Strain and Crystallographic Identification of the Helically Concaved Gap Surfaces of Chiral Nanoparticles." Nat Commun 14 (1)(2023).

5) Clark, J. N.; Huang, X.; Harder, R.; Robinson, I. K. "High-Resolution Three-Dimensional Partially Coherent Diffraction Imaging". Nat Commun, 3 (1) (2012).

6) L.E. Dresselhaus-Marais, B. Kozioziemski, T.S. Holstad, T.M. Ræder, M. Seaberg, D. Nam, S. Kim, S. Breckling, S. Choi, M. Chollet, P.K. Cook, E. Folsom, E. Galtier, A. Gonzalez, T. Gorkhover, S. Guillet, K. Haldrup, M. Howard, K. Katagiri, S. Kim, S. Kim, S. Kim, H. Kim, E.B. Knudsen, S. Kuschel, H.J. Lee,

C. Lin, R.S. McWilliams, B. Nagler, M.M. Nielsen, N. Ozaki, D. Pal, R. Pablo Pedro, A.M. Saunders, F. Schoofs, T. Sekine, H. Simons, T. van Driel, B. Wang, W. Yang, C. Yildirim, H.F. Poulsen, and J.H. Eggert, "Simultaneous bright- and dark-field X-ray microscopy at X-ray free electron lasers," Sci. Rep. **13**(1753), 1–19 (2023).

7) J.G. Ferrer, R. Rodríguez-Lamas, H. Payno, W. De Nolf, P. Cook, V.A.S. Jover, C. Yildirim, and C. Detlefs, "darfix – data analysis for dark-field X-ray microscopy," J. Synchrotron Radiat. **30**, 527–537 (2023)

8) Gonzalez, R.; Woods, R. "Digital Image Processing" (Addison-Wesley, 1993).

9) Gradwohl, K.-P.; Danilewsky, A. N.; Roder, M.; Schmidbauer, M.; Janicskó-Csáthy, J.; Gybin, A.; Abrosimov, N.; Sumathi, R. R. "Dynamical X-Ray Diffraction Imaging of Voids in Dislocation-Free High-Purity Germanium Single Crystals." J Appl Crystallogr, 53 (4), (2020).

10) Gaukroger, M.P.; Martineau, P.M.; Crowder, M.J.; Friel, I.; Williams, S.D.; Twitchen, D.J. "X-ray topography studies of dislocations in single crystal CVD diamond", Diam. Relat. Mater. 17(3), 262–269 (2008).

11) Han, T.R.T.; Zhou, F.; Malliakas, C.D.; Duxbury, P.M.; Mahanti, S.D.; Kanatzidis, M.G.; Ruan, C.Y. "Exploration of metastability and hidden phases in correlated electron crystals visualized by femtosecond optical doping and electron crystallography", Sci. Adv. 1(5) (2015).

12) Holstad, T. S.; Ræder, T. M.; Carlsen, M.; Bergbäck Knudsen, E.; Dresselhaus-Marais, L.; Haldrup, K.; Simons, H.; Nielsen, M. M.; Poulsen, H. F. "X-Ray Free-Electron Laser Based Dark-Field X-Ray Microscopy: A Simulation-Based Study". J Appl Crystallogr, 55 (1), (2022).

13) Holstad, T.S.; Dresselhaus-Marais, L.E.; Ræder, T.M.; Kozioziemski, B.; van Driel, T.; Seaberg, M.; Folsom, E.; Eggert, J.H.; Knudsen, E.B.; Nielsen, M.M.; Simons, H.; Poulsen, H.F. "Real-time imaging of acoustic waves in bulk materials with X-ray microscopy", Proc. Natl. Acad. Sci. 120(39), 1–4 (2023).

14) P.H. Huang, R. Coffee, and L. Dresselhaus-Marais, "Automatic Determination of the Weak-Beam Condition in Dark Field X-ray Microscopy", Integr. Mater. Manuf. Innov. **12**(2), 83–91 (2023).

15) Jakobsen, A. C.; Simons, H.; Ludwig, W.; Yildirim, C.; Leemreize, H.; Porz, L.; Detlefs, C.; Poulsen, H. F. "Mapping of Individual Dislocations with Dark-Field X-Ray Microscopy". J Appl

Crystallogr, 52 (1),(2019).

16) Kato, N.; Usami, K.; Katagawa, T. "The X-Ray Diffraction Image of a Stacking Fault." Adv. x-ray anal., 10,(1966).

17) Kim, Y.G.; Song, J.H.; Hong, S.; Ahn, S.H. "Piezoelectric strain sensor with high sensitivity and high stretchability based on kirigami design cutting." NPJ Flex. Electron. 6(1), 1–8 (2022).

18) Kutsal, M.; Bernard, P.; Berruyer, G.; Cook, P.K.; Hino, R.; Jakobsen, A.C.; Ludwig, W.; Ormstrup, J.; Roth, T.; Simons, H.; Smets, K.; Sierra, J.X.; Wade, J.; Wattecamps, P.; Yildirim, C.; Poulsen, H.F.; Detlefs, C. "The ESRF dark-field x-ray microscope at ID06", IOP Conf. Ser. Mater. Sci. Eng. 580(1), (2019).

19) S. Liu, C. Grech, M. Guetg, S. Karabekyan, V. Kocharyan, N. Kujala, C. Lechner, T. Long, N. Mirian, W. Qin, S. Serkez, S. Tomin, J. Yan, S. Abeghyan, J. Anton, V. Blank, U. Boesenberg, F. Brinker, Y. Chen, W. Decking, X. Dong, S. Kearney, D. La Civita, A. Madsen, T. Maltezopoulos, A. Rodriguez-Fernandez, E. Saldin, L. Samoylova, M. Scholz, H. Sinn, V. Sleziona, D. Shu, T. Tanikawa, S. Terentiev, A. Trebushinin, T. Tschentscher, M. Vannoni, T. Wohlenberg, M. Yakopov, and G. Geloni, "Cascaded hard X-ray self-seeded free-electron laser at megahertz repetition rate," Nat. Photonics **17**(November), 984–991 (2023).

20) Lorensen, W.E.; Cline, H.E. "Marching Cubes: A high Resolution 3D Surface Construction Algorithm," Comput. Graph. (ACM). 21(4), 163–169 (1987).

21) A. Madsen, J. Hallmann, G. Ansaldi, T. Roth, W. Lu, C. Kim, U. Boesenberg, A. Zozulya, J. Moller, R. Shayduk, M. Scholz, A. Bartmann, A. Schmidt, I. Lobato, K. Sukharnikov, M. Reiser, K. Kazarian, and I. Petrov, "Materials Imaging and Dynamics (MID) instrument at the European X-ray Free-Electron Laser Facility," J. Synchrotron Radiat. **28**, 637–649 (2021).

22) Masuya, S.; Hanada, K.; Oshima, T.; Sumiya, H.; Kasu, M. "Formation of stacking fault and dislocation behavior during the high-temperature annealing of single-crystal HPHT diamond," Diam. Relat. Mater. 75(December 2016), 155–160 (2017).

23) Moore, M. "Imaging diamond with x-rays", J. Phys. Condens. Matter 21(364217), 1–15 (2009).

24) Moore, R. G.; Lee, W. S.; Kirchman, P. S.; Chuang, Y. D.; Kemper, A. F.; Trigo, M.; Patthey, L.; Lu, D. H.; Krupin, O.; Yi, M.; Reis, D. A.; Doering, D.; Denes, P.; Schlotter, W. F.; Turner, J. J.;

Hays, G.; Hering, P.; Benson, T.; Chu, J.-H.; Devereaux, T. P.; Fisher, I. R.; Hussain, Z.; Shen, Z.-X. "Ultrafast Resonant Soft X-Ray Diffraction Dynamics of the Charge Density Wave in TbTe3". Phys. Rev. B 93 (2), 024304. (2016).

25) Pezeril, T. "Laser generation and detection of ultrafast shear acoustic waves in solids and liquids," Opt. Laser Technol. 83, 177–188 (2016).

26) M. Pergament, M. Kellert, K. Kruse, J. Wang, G. Palmer, L. Wissmann, U. Wegner, M. Emons, and M. Lederer, "340W femtosecond burst-mode non-collinear optical parametric amplifier for the european XFEL pump-probe-laser," Adv. Solid State Lasers Conf., 4–6 (2015).

27) Poulsen, H.F.; Jakobsen, A.C.; Simons, H.; Ahl, S.R.; Cook, P.K.; Detlefs, C. "X-ray diffraction microscopy based on refractive optics", J. Appl. Crystallogr. 50(5), 1441–1456 (2017).

28) Poulsen, H.F.; Cook, P.K.; Leemreize, H.; Pedersen, A.F.; Yildirim, C.; Kutsal, M.; Jakobsen, A.C.; Trujillo, J.X.; Ormstrup, J.; Detlefs, C. "Reciprocal space mapping and strain scanning using X-ray diffraction microscopy", J. Appl. Crystallogr. 51(5), 1428–1436 (2018).

29) Poulsen, H.F.; Dresselhaus-Marais, L.E.; Carlsen, M.A.; Detlefs, C.; Winther, G. "Geometrical-optics formalism to model contrast in dark-field X-ray microscopy", J. Appl. Crystallogr. 54(2018), 1555–1571 (2021).

30) Rodriguez-Fernandez, A.; Diaz, A.; Iyer, A. H. S.; Verezhak, M.; Wakonig, K.; Colliander, M. H.; Carbone, D. "Imaging Ultrafast Dynamical Diffraction Wave Fronts in Strained Si with Coherent X Rays." *Phys. Rev. Lett.* **127** (15)(2021).

31) H. Simons, A. King, W. Ludwig, C. Detlefs, W. Pantleon, S. Schmidt, F. Stöhr, I. Snigireva, A. Snigirev, and H.F. Poulsen, "Dark-field X-ray microscopy for multiscale structural characterization", Nat. Commun. **6**, 1–6 (2015).

32) Simons, H., S.R. Ahl, H.F. Poulsen, and C. Detlefs, "Simulating and optimizing compound refractive lens-based X-ray microscopes," J. Synchrotron Radiat. **24**(2), 392–401 (2017).

33) Simons, H.; Poulsen, H. F.; Schmidbauer, M.; Rodek, A.; Lehmann, S.; Madsen, A. "Dark-field x-ray microscopy for multiscale structural characterization", Phys. Rev. Mater. 2(8), 083803 (2018).

34) A. Snigirev, V. Kohn, I. Snigireva, and B. Lengeler, "A compound refractive lens for focusing high-energy X-rays", Nature **384**(6604), 49–51 (1996).

35) Takagi, S. "A dynamical theory of diffraction for a distorted crystal", J. Phys. Soc. Jpn. 26(3), 1239–1253 (1969).

36) M. Trigo et. al, "Fourier-transform inelastic X-ray scattering from time-and momentum-dependent phonon–phonon correlations", Nature Physics, 9 (12), 790-794 (2013)

37) T. Tschentscher, C. Bressler, J. Grünert, A. Madsen, A. Mancuso, M. Meyer, A. Scherz, H. Sinn, and U. Zastrau, "Photon Beam Transport and Scientific Instruments at the European XFEL," Appl. Sci. **7**(6), 592 (2017).

38) C. Yildirim, H.F. Poulsen, G. Winther, C. Detlefs, P.H. Huang, and L.E. Dresselhaus-Marais, "Extensive 3D mapping of dislocation structures in bulk aluminum", Sci. Rep. **13**(3834), 1–11 (2023).

39) D. Zhu et. al, "Phonon spectroscopy with sub-meV resolution by femtosecond x-ray diffuse scattering", Physical Review B 92, 054303 (2015)

40) Zusin, D.; Iacocca, E.; Le Guyader, L.; Reid, A. H.; Schlotter, W. F.; Liu, T.-M.; Higley, D. J.; Coslovich, G.; Wandel, S. F.; Tengdin, P. M.; Patel, S. K. K.; Shabalin, A.; Hua, N.; Hrkac, S. B.; Nembach, H. T.; Shaw, J. M.; Montoya, S. A.; Blonsky, A.; Gentry, C.; Hoefer, M. A.; Murnane, M. M.; Kapteyn, H. C.; Fullerton, E. E.; Shpyrko, O.; Dürr, H. A.; Silva, T. J. Ultrafast Perturbation of Magnetic Domains by Optical Pumping in a Ferromagnetic Multilayer. Phys. Rev. B 2022, **106** (14), 144422.(2022)

# Supplemental information:
# Dark-Field X-ray Microscopy for 2D and 3D Imaging of Microstructural Dynamics at the European X-ray Free Electron Laser

## I) Scaling

For the laboratory coordinate system, we define the sample position such that the $x_\ell$-axis to be along the length of the sample, as shown in Fig. S1. The sample is translated along the $x_\ell$-axis during the "x-scan" described in the main text. The $z_\ell$-axis is along the width of the sample parallel to the surface normal. The $y_\ell$-axis points into the plane of the page in Fig S1. The origin of the $x_\ell$-$z_\ell$ plane in the coordinate system in the sample is the point where the incident line beam exists in the sample in Fig S1.

We also define a coordinate system at the two-dimensional detector with unit vectors $y_i$ along the height of the detector and $x_i$ into the paper as drawn in Fig S1. A similar scaling procedure is also described in Dresselhaus 2023, but due to the 3D structure scan further explanation was necessary in this work. [ Dresselhaus 2023]

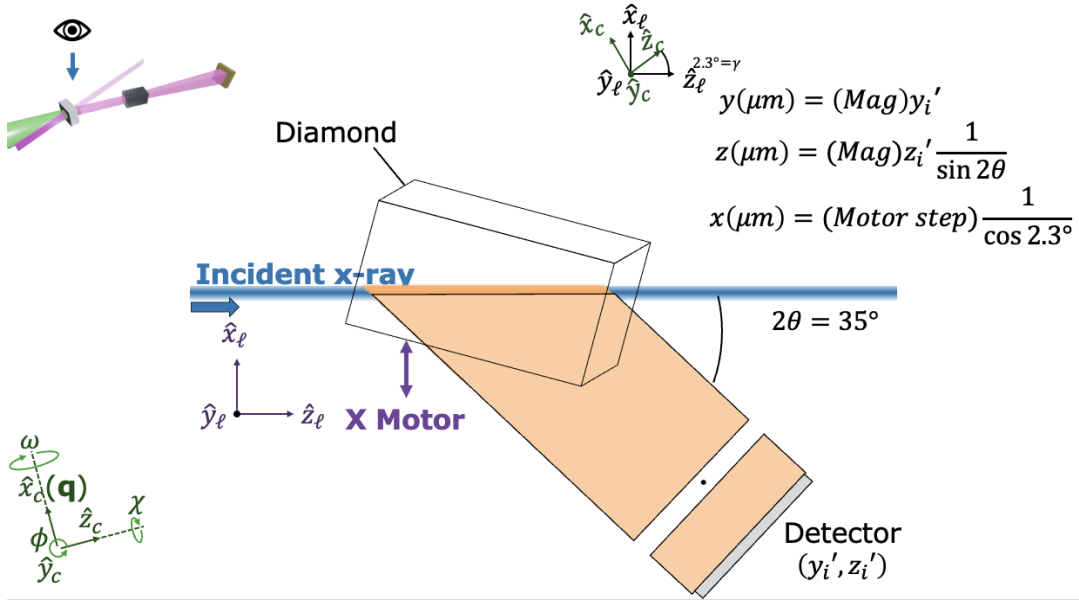

**Supplemental Fig 1:** Schematic of the experimental geometry and coordinate systems from the top perspective (as shown by the schematic in the upper left corner). The beam path and relationships between the *x, y,* and *z* demonstrate the image stretching that occurs along the diffracted beam as it changes the resulting image, as quantified by the equations. Image adapted from [Wang 2024]

## II) Data processing

First motor stability was checked and data was sorted based on the stable motor positions. Data in the stable motors positions was then averaged, and referred to in the code as a frame, to reduce the acquired data set from the initial ~3000 images to the most representative images per scan step. This data processing also involved subtracting the dark, we did take multiple darks throughout the experiment but observed no difference between dark frames; as such we used the same dark for all frames processed for ease of computation.

Center of mass (COM) images were created using the formalism published in [Ferrer 2023], though a gaussian filter was used instead of the median filter described in that work. The included data (see repository at the end of the supplement) is already processed as described above. The data for the energy scan included all of the $\varphi$ (motor alias: SSRY) motor positions for each scan in the array. Since the 3D structure scanned both $\varphi$ motor and $x$ (motor alias: BSMZ) the array attached is the selected $\varphi$ positions for every $x$ motor position.

## III) Energy Scan

**Gaussian Fit to ensure phi motor stability**

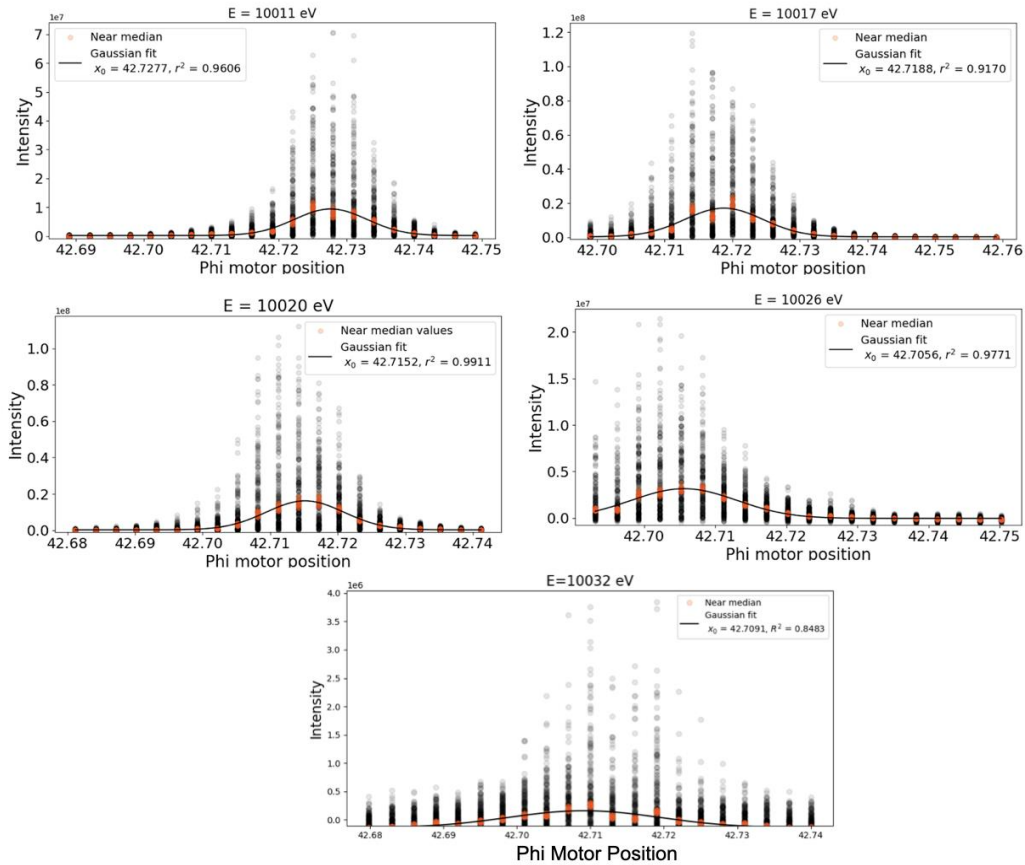

**Supplementary Fig 2,** Summed rocking curve of all scans collected at each photon energy. Energy 10032eV was removed from the rest of the analysis.

The intensity vs $\varphi$ positions for each photon energy in the energy scan was fit to a Gaussian to evaluate its COM $\varphi$ position. Energy 10032eV was omitted from the COM reconstruction in Figure 4, due to the much worse gaussian fit and the 100x lower intensity than the strong-beam conditions measured at 10020 eV.

Our first methodology for choosing the weak-beam positions for each $E_n$ position was to select the same offset in standard deviations based on the Gaussian fit. As shown in in Fig 2, due to the limited data points, this option was not as accurate especially for the 10026eV energy for which the $\varphi_{SB}$ was at the edge of the scan. Therefore we decided to use the same $\varphi$ motor position instead of the normal statistical standard for more reproducible results.

### IV) Energy Scan code

The Energy Scan code includes all of the preprocessed full $\varphi$ vs energy scans. The energy scan position is chosen as described above. Due to the change in energies we predict that this created a slight change in the field of view, as the surface would move for different energies. As such we corrected this by assuming that the surface strain would be in the same position regardless of energy and aligning to that. We note that this was a proof of concept scan and further adjustments need to be made in the methodology in order to create a more accurate scanning method.

### IV) 3D Structure code

The code for the XFEL 3D structure scans uses pre-processed and pre-selected images acquired at the weak-beam conditions for each $x$-layer, as the whole data set due to file size constraints. As such, here we describe the methodology for choosing the weak-beam frame for each of the 40 steps in the $x$-$\varphi$ scan.

The weak-beam frames were first selected for images with a $\varphi$ position at 3σ from the center of the Gaussian fit of intensity vs $\varphi$ scans of each layer. A $\varphi$ motor value was then selected that was the closest to the $\varphi = 3σ$ value, which was ultimately too dim for image clarity. We then manually adjusted the $\varphi$ position for all layers to be 1 step earlier than the 3σ value. The $\varphi$ position that exhibited the weak-beam contrast with the highest image clarity was then selected manually based on the nearby $\varphi$ values. The attached data includes the selected weak beam, full data is available upon request.

**Supplementary Citations:**
[1] Y. Wang, N. Bertin, D. Pal, S. J. Irvine, K. Katagiri, R. E. Rudd, L. E. Dresselhaus-Marais. "Computing virtual dark-field X-ray microscopy images of complex discrete dislocation structures from large-scale molecular dynamics simulations" In Review. *arXiv:2409.01439* (2024).

[2] L.E. Dresselhaus-Marais, B. Kozioziemski, T.S. Holstad, T.M. Ræder, M. Seaberg, D. Nam, S. Kim, S. Breckling, S. Choi, M. Chollet, P.K. Cook, E. Folsom, E. Galtier, A. Gonzalez, T. Gorkhover, S. Guillet, K. Haldrup, M. Howard, K. Katagiri, S. Kim, S. Kim, S. Kim, H. Kim, E.B. Knudsen, S. Kuschel, H.J. Lee, C. Lin, R.S. McWilliams, B. Nagler, M.M. Nielsen, N. Ozaki, D. Pal, R. Pablo Pedro, A.M. Saunders, F. Schoofs, T. Sekine, H. Simons, T. van Driel, B. Wang, W. Yang, C. Yildirim, H.F. Poulsen, and J.H. Eggert, "Simultaneous bright- and dark-field X-ray microscopy at X-ray free electron lasers," *Sci. Rep.* **13**(1753), 1–19 (2023).

[3] J.G. Ferrer, R. Rodríguez-Lamas, H. Payno, W. De Nolf, P. Cook, V.A.S. Jover, C. Yildirim, and C. Detlefs, "darfix – data analysis for dark-field X-ray microscopy," *J. Synchrotron Radiat.* **30**, 527–537 (2023)